\documentclass[aps,10pt,pra,floatfix,
reprint,
superscriptaddress,%
]%
{revtex4-1}

\usepackage{bm}
\usepackage{amsmath,amssymb}
\usepackage{comment}
\usepackage{bm}
\usepackage{braket}
\usepackage{braket}
\usepackage[dvips]{graphicx}
\usepackage{color}
\usepackage[colorlinks=true]{hyperref}

\hypersetup{
    colorlinks=true,
    linkcolor=blue,
    filecolor=blue,      
    urlcolor=blue,
}

\usepackage[dvipsnames]{xcolor}

\newcommand{\B}{\mathbf}
\newcommand{\pvec}[1]{\vec{#1}\mkern2mu\vphantom{#1}}
\begin{document}

\title{Topological superconductivity in proximity to type-II superconductors}

\author{Alexander Nikolaenko}
\affiliation{
Karazin Kharkiv National University, Kharkiv 61022, Ukraine
}
\affiliation{
Max Planck Institute for the Physics of Complex Systems, N\"othnitzer Str. 38, 01187 Dresden, Germany
}
\author{Falko Pientka}%
 \email{pientka@itp.uni-frankfurt.de}
\affiliation{
Institut f\"ur Theoretische Physik, Goethe-Universit\"at, 60438 Frankfurt am Main, Germany
}
\affiliation{
Max Planck Institute for the Physics of Complex Systems, N\"othnitzer Str. 38, 01187 Dresden, Germany
}
\date{\today}

\begin{abstract}    

One-dimensional systems proximity-coupled to a superconductor can be driven into a topological superconducting phase by an external magnetic field. Here, we investigate the effect of vortices created by the magnetic field in a type-II superconductor providing the proximity effect. We identify different ways in which the topological protection of Majorana modes can be compromised and discuss strategies to circumvent these detrimental effects. Our findings are also relevant to topological phases of proximitized quantum Hall edge states. 
\end{abstract}

\maketitle

{\em Introduction.}---
Topological superconductors are thought to exist in numerous different platforms, ranging from one-dimensional materials such as atomic chains \cite{Nadj-Perge2014,Ruby2015a} and semiconductor wires \cite{Oreg2010,Lutchyn2010,Mourik2012,Albrecht2016} to emergent one-dimensional systems such as edge modes of quantum Hall systems \cite{Lindner2012,Clarke2013,Lee2017} and two-dimensional topological insulators \cite{Fu2009,Bocquillon2017,Deacon2017} or Josephson junctions \cite{Pientka2017,Ren2019,Fornieri2019}. The great variety of proposals originates in the conceptual simplicity of topological superconductors: any one-dimensional spinless superconductor is topological. 
While pairing can be reliably be induced in one-dimensional systems via the proximity effect of a parent $s$-wave superconductor \cite{Franceschi2010}, the technological challenge is to break the spin degeneracy in a controlled way that does not impair the induced superconductivity. Experimental progress so far has mostly been based on thin superconducting films that can withstand moderate in-plane magnetic fields \cite{Albrecht2016}. 

\begin{figure}[tb]
  \center{\includegraphics[width=1\linewidth]{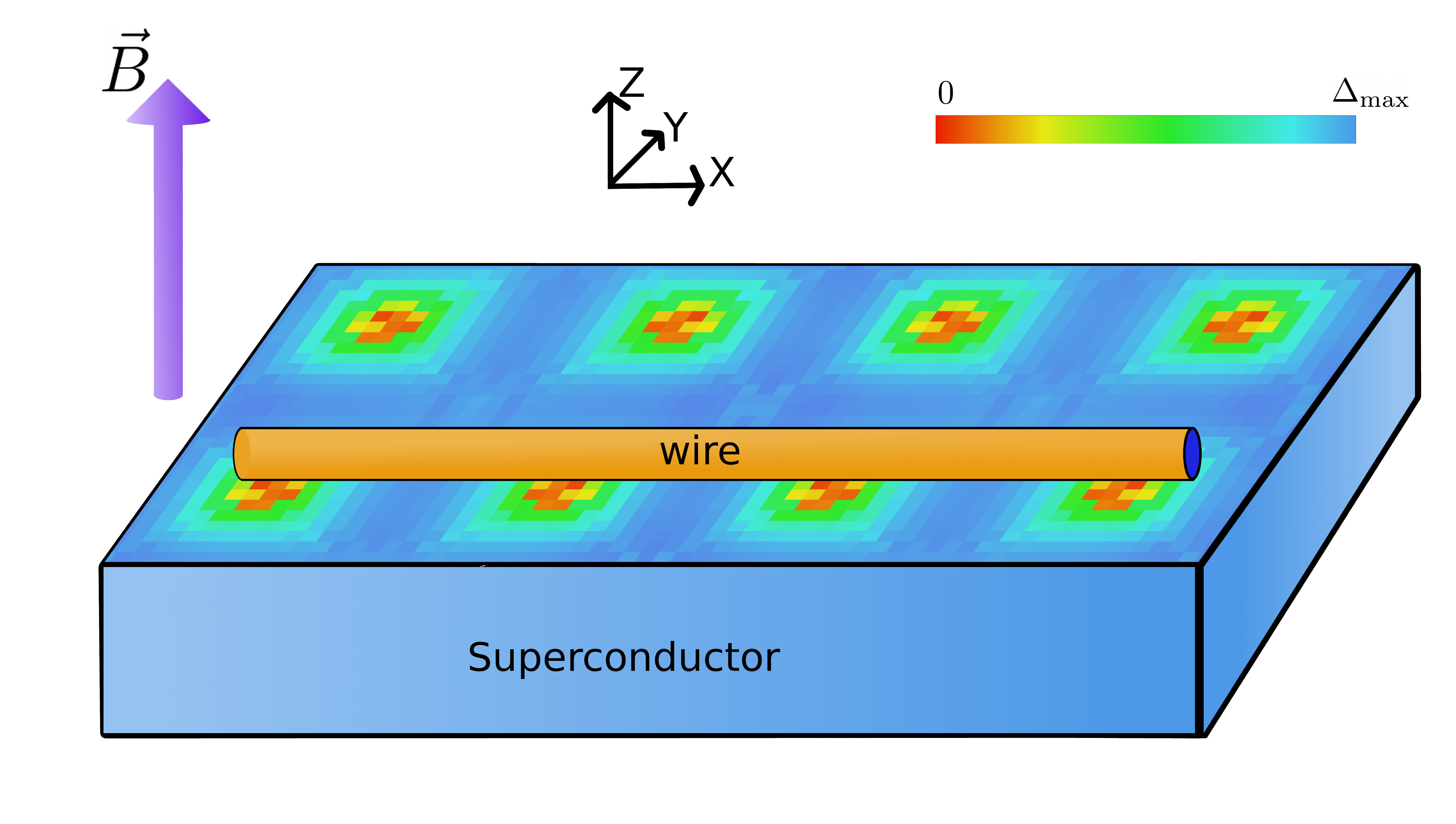}}
  \center{\includegraphics[width=1\linewidth]{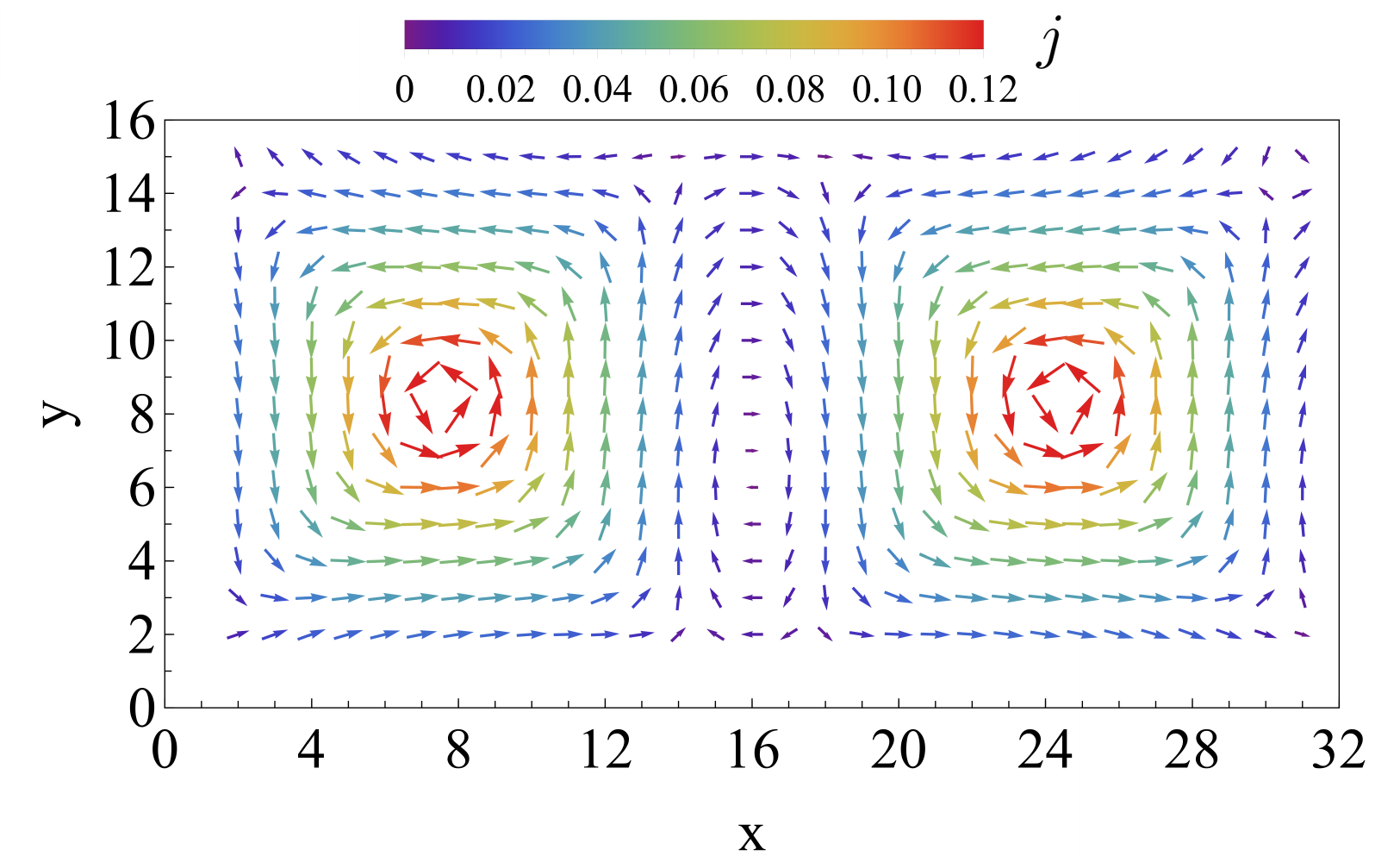}}
\llap{\parbox[c]{1.0cm}{\vspace{-20.8cm}\hspace{-7.2cm}(a)}}
\llap{\parbox[c]{1.0cm}{\vspace{-10.8cm}\hspace{-7.2cm}(b)}}
  \caption{(a) A wire is placed on top of a superconductor in a perpendicular magnetic field. The magnetic field creates vortices, where the superconducting order parameter is suppressed as shown by the color scale on the top surface of the superconductor. 
 (b) The vortex cores are encircled by supercurrents, whose magnitude increases towards the vortex core. The interaction is chosen to be $V=-1.95w^s$. The size of the magnetic unit cell  is $N_x\times N_y=32\times 16$. The number of magnetic unit cells in each direction is $M_x=16$ and $M_y=1$ although this choice does not affect the results strongly. }
  \label{fig:gradient}
\end{figure}

An alternative avenue are type-II superconductors which sustain sizable magnetic fields even in the bulk by allowing for vortices. They are indispensable for topological superconductors platforms based on the quantum Hall effect \cite{Lee2017} and, in particular, for realizations of parafermions in proximity-coupled fractional quantum Hall systems \cite{Lindner2012,Clarke2013}. Type-II superconductors can even be beneficial for nanowire realizations as they allow for out-of-plane magnetic fields, which enables more flexible designs of nanowire networks needed for topological quantum computation \cite{Karzig2017}. Here, we elucidate the effect of vortices inside an $s$-wave superconductor on the induced topological phase in a proximity coupled one-dimensional system [see setup in Fig.~\ref{fig:gradient}(a)]. We focus, in particular, on the decay length of Majorana end states as a measure of the topological protection.

{\em Theoretical model.}---Our starting point is the Hamiltonian $H=H^d+H^s+H^t$, where 
\begin{align}
H^d=&\sum_{l,\sigma}(\epsilon_d-\mu)d^{\dag}_{l,\sigma}d_{l,\sigma}-\sum_{l,\sigma}(w_{l,l+1}d^{\dag}_{l,\sigma}d_{l+1,\sigma}+{\rm h.c.})\notag\\
&-K\sum_{l,\sigma,\sigma^{'}} \bm{S}_l d^{\dag}_{l,\sigma} \bm{\sigma}_{\sigma,\sigma'} d_{l,\sigma'}
\end{align}
describes a one-dimensional chain with on-site energy $\epsilon_d$, chemical potential $\mu$ and nearest-neighbor hopping strength $w$. The operator $d_l$ annihilates a fermion in the chain at site $\B R_l=(x_l,y_c)$, where $y_c$ is fixed and $l$ runs from 1 to $L_c$.
We assume the wire to have helical magnetic order with an exchange splitting $K$ and a spin texture $\bm{S}_l=S(\cos \alpha_l,\sin \alpha_l,0)$ and $\alpha_l=2k_h x_l$.
The chain is placed on top of a two-dimensional superconductor with nearest-neighbor hopping of strength $w^s$ and attractive interaction of strength $V$ described by the Hamiltonian
\begin{align}
 H^s=&-\sum_{i,\sigma}\mu c^{\dag}_{i,\sigma}c_{i,\sigma}-\sum_{\braket{i,j},\sigma}w^s_{ij}c^{\dag}_{i,\sigma}c_{j,\sigma}\notag\\
&+V\sum_{i} c^{\dag}_{i\uparrow}c_{i \uparrow}c^{\dag}_{i\downarrow}c_{i \downarrow}.
\end{align}
The operator $c_{j}$ annihilates a fermion in the superconductor at site $\B R_i=(x_i,y_i)$ defined on a square lattice of size $L_x\times L_y$. The superconductor and the chain are coupled via tunneling at sites $\B R_l$
\begin{align}
H^t=-t \sum_{l,\sigma}(c^{\dag}_{l,\sigma} d_{l,\sigma}+d^{\dag}_{l,\sigma} c_{l,\sigma}).
\end{align}
In the presence of a magnetic field perpendicular to the superconductor, the hopping amplitudes acquire Peierls phases,
$w^s_{ij}= w^s\exp(2\pi i \int^{\bm{R_j}}_{\bm{R_i}}\vec{A} \cdot d\vec{r}/\Phi_0)$, where $\Phi_0=h/e$ is the flux quantum and a similar relation for $w_{l,l+1}$. We assume periodic boundary conditions for the superconductor and open boundary conditions for the chain unless stated otherwise. 
Throughout the paper we chose $w^s=1$, $\mu=0.5$, $K=1$, $S=2$, and $\epsilon_d=3$.

While the Hamiltonian describes a wire with a spin helix, it can be mapped to a ferromagnetic wire (or a wire with a large Zeeman splitting due to an external field) in proximity to a superconductor with spin-orbit coupling. This can be seen by rotating all spins onto a single axis and performing the unitary transformation 
$d_{j,\sigma} \rightarrow \exp(-i \alpha_j/2 \sigma_z)d_{j,\sigma}$, $c_{j,\sigma} \rightarrow \exp(-i \alpha_j/2 \sigma_z)c_{j,\sigma}$. As a result, the chain fermions have spins aligned along $x$ and the hopping amplitudes in the superconductor transform as $w^s_{ij}\rightarrow \exp(i  k_h (x_i-x_j) \sigma_z)w_{ij}$  (and equivalently for $w_{ij}$), which couples the spin to orbital motion along the $x$ direction. We do not expect qualitative changes when a more realistic type of spin-orbit coupling is assumed because the topological phase is predominantly affected by the spin-orbit component along the chain.

We approach the problem in several steps. We first determine the gap in the isolated superconductor $H^s$ in a magnetic field self-consistently within mean-field theory. From this we can obtain the superconductor Green's function in real space. Finally, we can obtain the spectrum and the Majorana wavefunction from the Green's function of the full system that accounts for the coupling between chain and superconductor via a Dyson equation. Notice that this approach ignores a possible suppression of the superconducting gap due to the coupling to the chain, which is valid in the weak-coupling limit.

The superconducting Hamiltonian $H^s$ can be described in mean-field theory by the Bogoliubov-de Gennes equation \cite{deGennes1994}
\begin{gather}
{\cal H}^s\begin{pmatrix}
u_n\\
v_n
\end{pmatrix}=
\begin{pmatrix}
h & \Delta \\
\Delta^* & -h^*
\end{pmatrix}
\begin{pmatrix}
u_n  \\
v_n
\end{pmatrix}
=E_n
\begin{pmatrix}
u_n\\
v_n
\end{pmatrix}.
\end{gather}
The Hamiltonian is spin degenerate and we can therefore suppress the spin indices. The Nambu vector $(u, v)^T$ has dimension $2 L_x L_y$, $\Delta$ is 
a diagonal matrix, and $h_{ij}=-\delta_{ij}\mu-w^s_{ij}$.
The lattice translation symmetry is broken by the magnetic field, however, one can construct magnetic translation operators that commute with the Hamiltonian \cite{Bernevig2013}.
The magnetic unit cell is larger by a factor $N_x \times N_y$, where $N_x N_y a^2 B = \Phi_0$, and the total size of the lattice is chosen to be multiple of the unit cell $L_x\times L_y=M_x N_x \times M_y N_y$. The magnetic translation operators give rise to a magnetic Bloch theorem
\begin{align}
\begin{pmatrix}
u_{n,k}(\vec{r}+\vec{R}) \\
v_{n,k}(\vec{r}+\vec{R})
\end{pmatrix}
=e^{i \vec{k} \vec{R}}
\begin{pmatrix}
e^{i\chi(\vec{r},\vec{R})/2}u_{n,k}(\vec{r})\\
e^{-i\chi(\vec{r},\vec{R})/2}v_{n,k}(\vec{r})
\end{pmatrix},
\end{align}
where $\vec{R}=m N_x \vec{e}_x+n N_y \vec{e}_y$ is a unit cell vector and $\chi(\vec{r},\vec{R})=-4\pi/\Phi_0 \vec{A}(\vec{R})\cdot \vec{r}=4\pi n \, x/N_x$ in the gauge $\vec{A}=-B y \vec{e}_x $. Hence, the self-consistency equation takes the form (see App.~\ref{Appendix_A}) \cite{Zhu1995}
\begin{align}
\Delta(\vec{r}+\vec{R})=\frac{V}{M_x M_y} e^{i\chi(\vec{r},\vec{R})}\sum_{n,k}u_{n,k}(\vec{r})v_{n,k}^*(\vec{r})f(E_{n,k})
\label{delta}
\end{align} 
with $f$ the Fermi distribution.

We now return to the full system of the chain coupled to the superconductor. The Green's function of the chain satisfies the Dyson equation
\begin{align}
g^{d}=(1-\Sigma g^{d}_{0})^{-1}g^{d}_{0}
\end{align}
where the self-energy $\Sigma={\cal T} g^{s}_{0}{\cal T}$ describes the tunneling to the superconductor. Here  $g^{d}_{0}=(E-\mathcal{H}_d)^{-1}$ and $g^{s}_{0}=(E-\mathcal{H}_s)^{-1}$ are the real-space Nambu Green's functions of the chain and superconductor in the absence of a coupling and ${\cal T}$ is the tunneling matrix, which equals $ t\tau_z$ on the sites covered by the chain and zero otherwise.
The spectrum corresponds to poles of the Green's function and can hence be obtained from  
\begin{align}
\text{Det}({1-{\cal T}g^{s}_{0}{\cal T}g^{d}_{0}})=0.
\label{spectrum_e}
\end{align}
Notice that ${\cal T}g^{s}_{0}{\cal T}$ depends only on the superconductor Green's function at the lattice sites covered by the chain. To find the wavefunction of zero modes, it is therefore sufficient to consider a reduced Green function $\tilde g^s_0$, which is given by $g^s_0$ projected to the chain sites \cite{Peng2015}. The wavefunction of a zero mode is then given by the kernel of
\begin{gather}
g^{-1}=
\begin{pmatrix}
(\tilde{g}^{ss}_0)^{-1} & \ t \tau_z\\
t \tau_z & (g^{dd}_0)^{-1}
\end{pmatrix}
\label{wave}
\end{gather}
evaluated at zero energy. The local density of states (LDOS) is simply  $n(r,E)=-(1/\pi){\rm Im}g_{ee}(E,r)$, where $g_{ee}$ is the electronic part of the Green's function.

{\em Topological phase.}---
The perpendicular magnetic field introduces vortices into the superconductor. To obtain the spatial dependence of the superconducting order parameter, we solve the self-consistency equation~(\ref{delta}). The topological phase in the chain can be affected both by the suppression of the pairing strength as well as by the gradient of the superconducting phase due to supercurrents encircling the vortices \cite{Romito2012}. The phase gradient between neighboring sites $i$ and $j$ can be written in a gauge invariant form as
\begin{align}
\Delta \theta_{i,j}=\phi_{j}-\phi_{i}+\frac{4\pi}{\Phi_0}\int_{\B R_i}^{\B R_j} d\B r\cdot \B A(\B R_{i})\label{gauge_inv_gradient}
\end{align}
where $\phi_i=\arg\Delta_i$. The supercurrent flowing between sites $i$ and $j$ is related to the phase gradient by  $j_{i,j}=(2en_s/m)\sin \Delta \theta_{i,j}$, where $n_s$ is the geometric mean of the condensate density at the two sites.
Figure~\ref{fig:gradient}(b) shows the real-space image of the vectorial supercurrent at each site in the magnetic unit cell. As each vortex carries a flux $h/2e$, there are two vortices in each unit cell. The supercurrent decreases further away from the vortex cores and it essentially zero half-way between two vortices.

\begin{figure}[tb]
\includegraphics[width=1\linewidth]{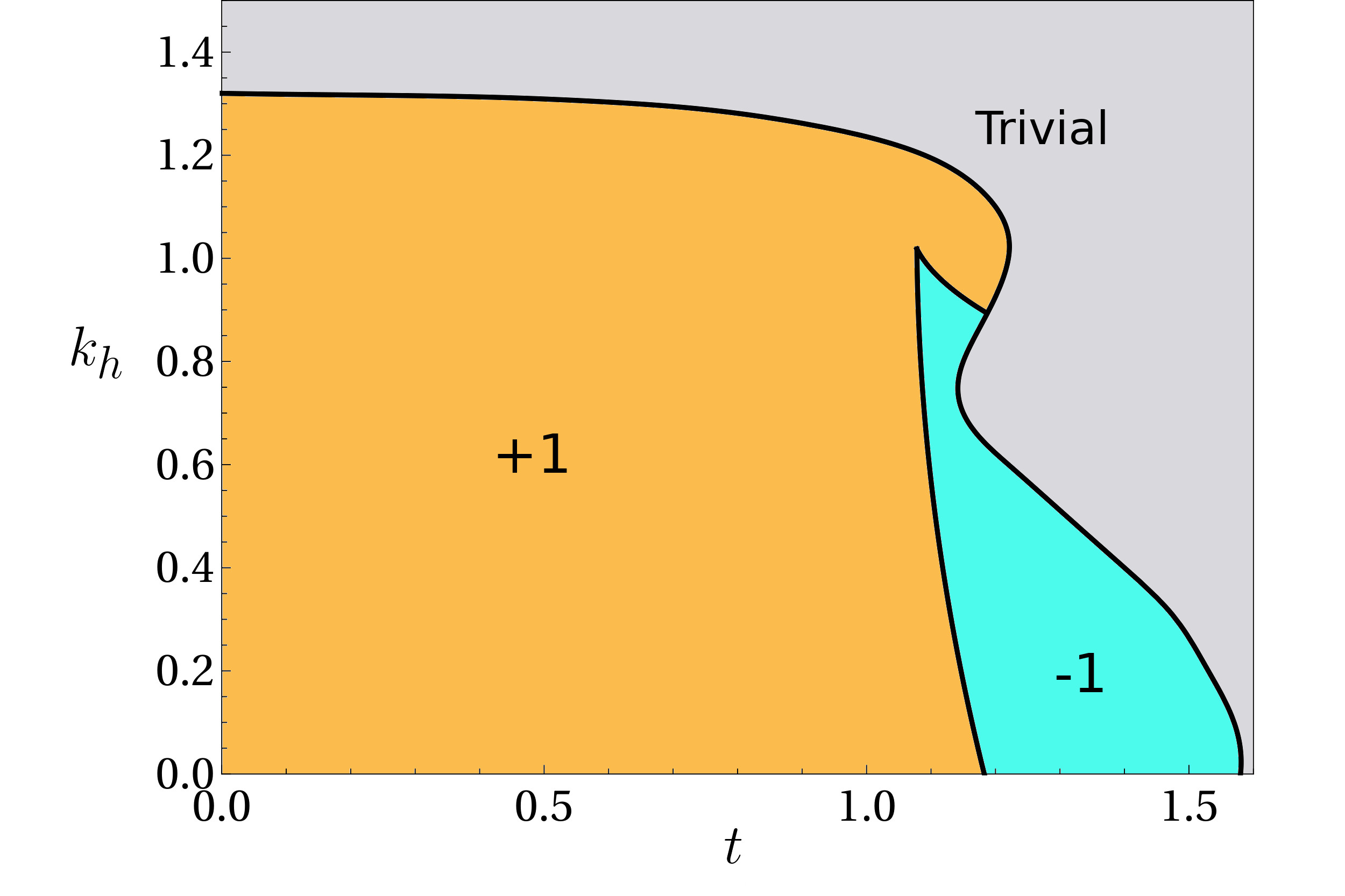}
  \includegraphics[width=1\linewidth]{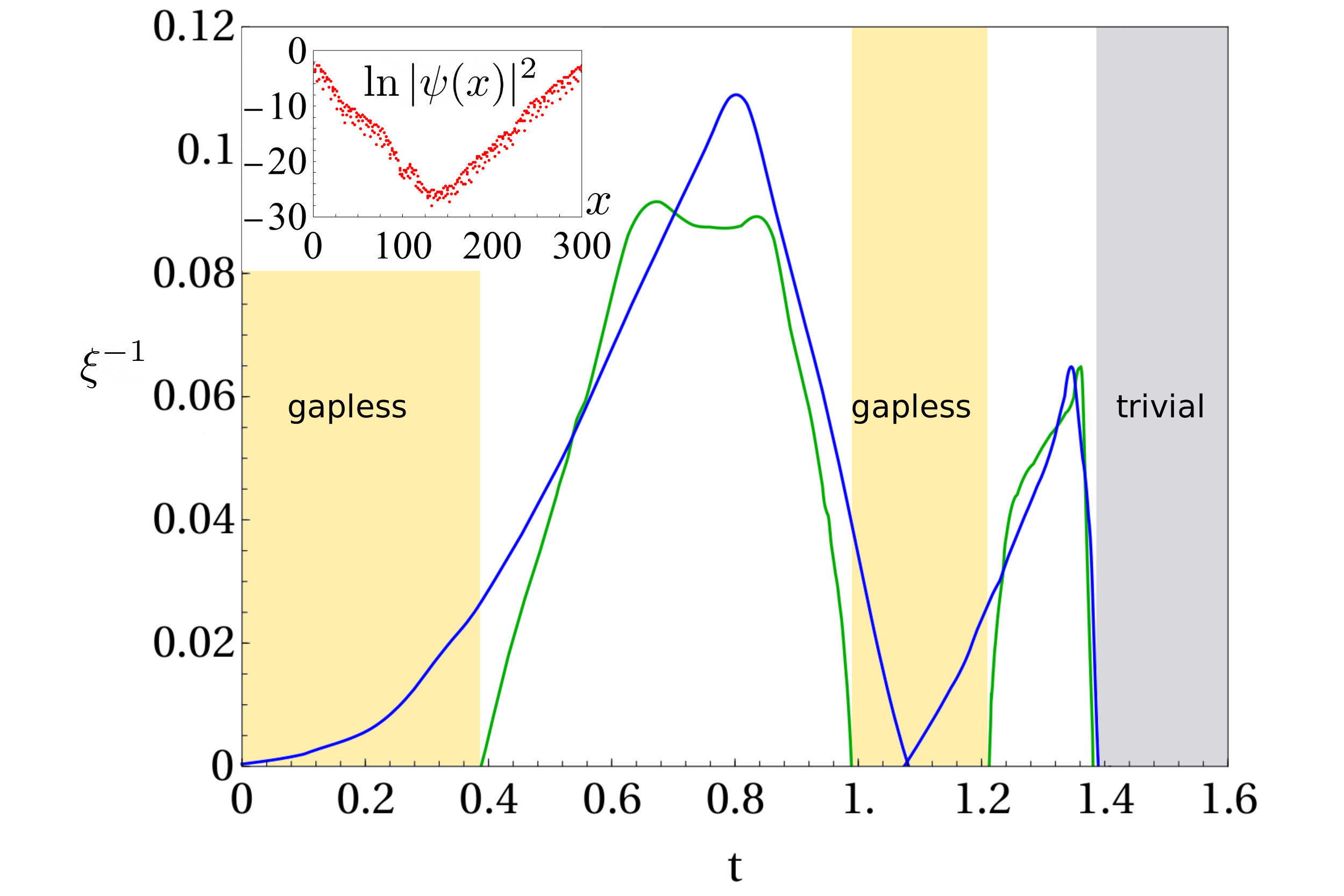}
\llap{\parbox[c]{1.0cm}{\vspace{-23.2cm}\hspace{-7.2cm}(a)}}
\llap{\parbox[c]{1.0cm}{\vspace{-11.5cm}\hspace{-7.2cm}(b)}}
  \caption{(a) Phase diagram for the wire coupled to a homogeneous superconductor. The gray area corresponds to the trivial phase, yellow to the topological phase with $\mathbb{Z}=1$ and cyan to the topological phase with $\mathbb{Z}=-1$. We chose $V=-1.95$ and $N_y=64$. The wire and the superconductor are periodic in $x$-direction. (b) Dependence of the Majorana coherence length on the vertical hopping $t$ between the wire and the superconductor. The blue line corresponds to a homogeneous superconductor and the green line to the superconductor with vortices. We chose $V=-1.95$, $k_h=0.4$, and $y_c=2$. The sizes of the systems are: $N_x\times N_y=32\times 16$, $M_x\times M_y=16\times 1$, $L=300$. The inset shows a semi logarithmic plot of the Majorana wavefunction  for the case of a   superconductor with vortices and $t=0.6$.}
  \label{fig:top}
\end{figure}

Before we consider the effect of vortices on the topological phase, we first consider the case of zero orbital magnetic field as a reference. In this case, the Hamiltonian possess both particle-hole and time-reversal symmetry there are different topological phases characterized by a $\mathbb{Z}$ number \cite{Tewari2012}. The corresponding phase diagram calculated for a chain with periodic boundary conditions is shown in Fig.~\ref{fig:top}(a) (see App.~\ref{App_B}).
The parameters are chosen such that the chain Hamiltonian ${\cal H}_d$ has a single band that crosses zero energy. Accordingly, at small coupling to the superconductor, the chain is typically in the topological phase. At strong coupling $t\gtrsim w^s$ the system eventually becomes trivial. This can be understood as follows: in the limit of strong coupling, the chain fermions $d_l$ on each site dimerize with the neighboring fermions in the superconductor $c_l$, pushing the spin-polarized states away from the Fermi level. As the model has a chiral symmetry in the absence of vortices, another phase transition occurs inside the topological phase between phases with topological index $\nu=\pm1$ in agreement with previous studies on a related model \cite{Peng2015}. 

The Majorana states are topologically protected by their spatial separation. The scale of separation, below which the Majoranas gap out and no longer form true zero modes, is given by the induced coherence length, which can therefore serves as a measure of the topological protection. The coherence length can be determined by fitting the real-space exponential decay of the Majorana wavefunction obtained from the kernel of Eq.~(\ref{wave}) [see inset of Fig.~\ref{fig:top}(b)]. Figure~\ref{fig:top}(b) compares the inverse Majorana decay length for a superconductor with and without vortices. 
At zero coupling, the induced gap vanishes and the coherence length diverges accordingly. At larger coupling strengths, the phase transition between the two topological phases with $\nu=\pm 1$ in the absence of vortices is signaled by gap closing at a critical point. An out-of-plane magnetic field results in supercurrents, which break the chiral symmetry such that the topological index reduces to a $\mathbb{Z}_2$ number (see App.~\ref{App_B}). The critical point is broadened into a gapless phase (cf.~Refs.~\onlinecite{Romito2012} and \onlinecite{Pientka2013a}). 

\begin{figure}[tb]
  \center{\includegraphics[width=.97\linewidth]{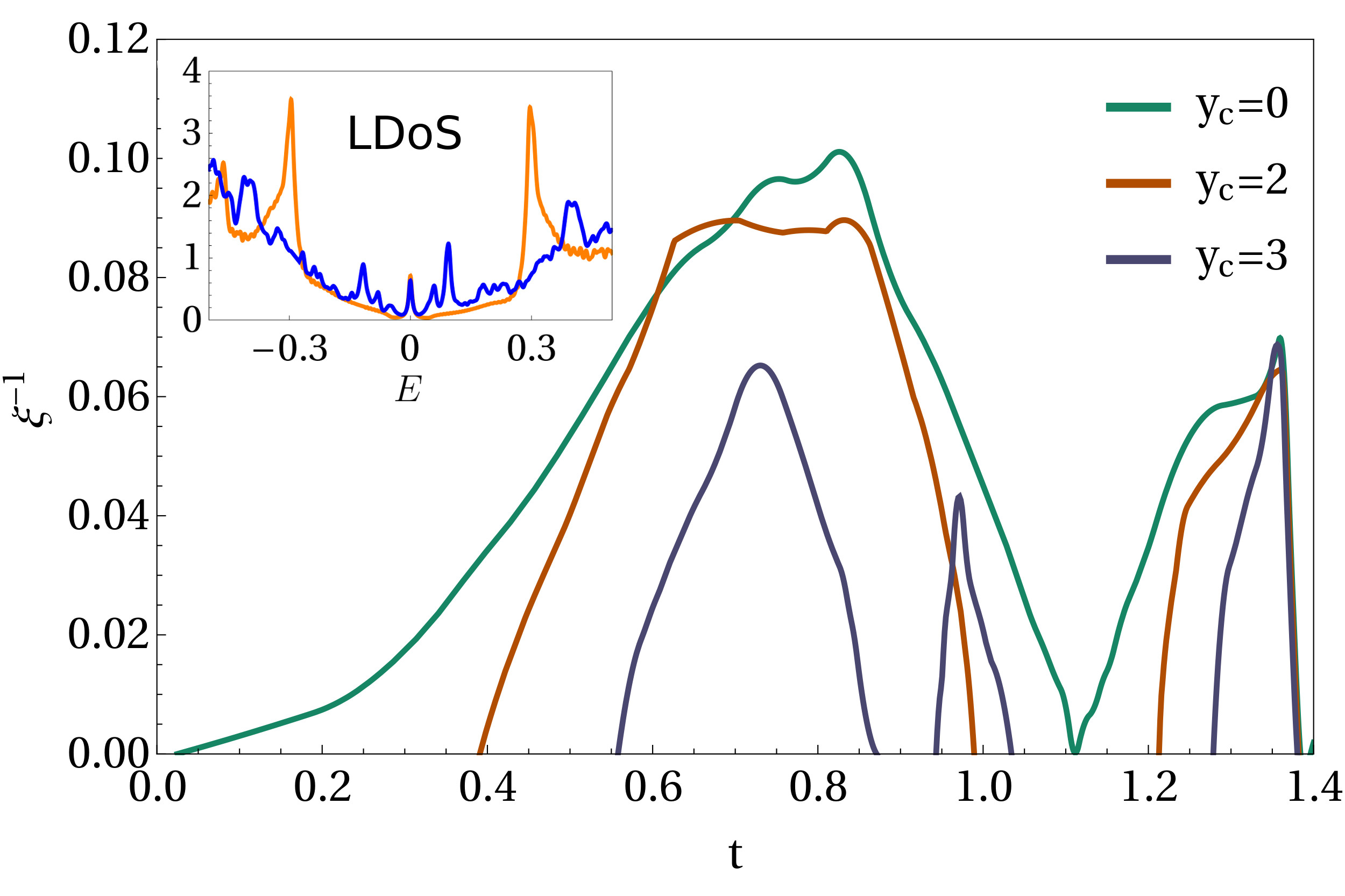}}

  \caption{The coherence length for different locations of the wire relative to the vortex cores inside the superconductor, where $y_c=0$ refers to the position half-way between two vortices.   The inset shows the LDOS at site $x=1$, $y_c=1$ for a homogeneous superconductor (orange line) and a superconductor with vortices (blue line) with $t=0.5$. Other parameters are chosen to be the same as in Fig.~\ref{fig:top}(b). }
  \label{fig:coherence}
\end{figure}

Placing the wire closer to the vortices increases the maximal phase gradient along the wire as can be seen from Fig.~\ref{fig:gradient}(b). As a result the size of the gapless phase grows as the wire approaches the vortices. This trend is visible in Fig.~\ref{fig:coherence}, which compares the inverse coherence length for different positions of the wire. Besides the reduced topological phase space, the protection of the Majorana modes inside the topological phase is decreased as the induced coherence length increases. Importantly, however, the phase gradient is basically zero when the wire is right between two vortices, $y_c=0$, [see Fig.~\ref{fig:gradient}(a)] and hence the corresponding phase diagram and induced coherence length is similar to the case without vortices. 

In addition to introducing phase gradients, vortices can degrade the topological protection of Majorana states by reducing the superconducting gap. In a tunneling experiment measuring the LDOS \cite{Ruby2015}, this results in an enhanced spectral weight at low energies. The inset of Fig.~\ref{fig:coherence} compares the LDOS in the topological phase with and without vortices. While the zero-energy peak remains largely unaffected by the magnetic flux, the finite bias signatures are changed drastically. Most importantly, the coherence peaks of the superconducting substrate at $E\simeq 0.3$ are suppressed in the case with vortices. The corresponding spectral weight is partially transferred to lower lying Caroli-de Gennes-Matricon states localized inside the vortices, which appear as a series of low-energy peaks in the LDOS.

 \begin{figure}[tb]
  \center{\includegraphics[width=.97\linewidth]{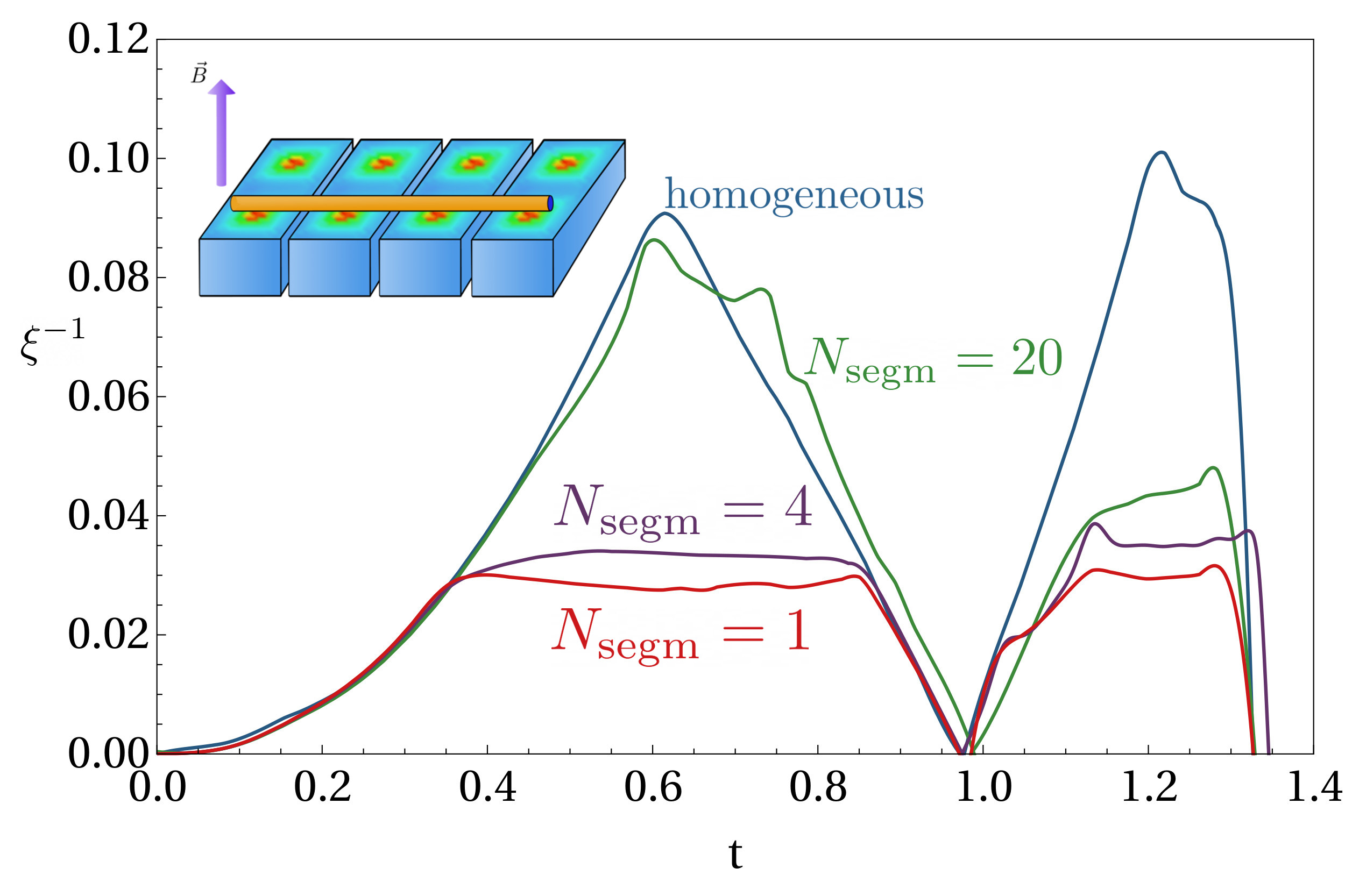}}
   
  \caption{ The inverse Majorana decay length when the superconductor is divided into several segments. The plot compares a homogeneous superconductor without vortices (blue line), with a superconductor with vortices in one segment (red line), four segments (purple line) and twenty segments (green line). The
  inset shows a schematic representation of the wire above a superconductor with several isolated parts. The parameters are $y_c=0$, $V=-1.6$, $k_h=0.4$, $N_x\times N_y=64 \times 32$, $M_x \times M_y=20\times16$, and $L=900$.}
  \label{fig:divided}
  \end{figure}

The presence of bound vortex states can have a dramatic effect on the protection of Majorana states. Figure~\ref{fig:divided} compares the Majorana decay length with (red line) and without (blue line) vortices at $y_c=0$, where the phase gradient is essentially absent and the suppression of the superconducting order parameter is minimal.
In the vicinity of the phase transitions, where the coherence length is long, the presence of vortices has negligible effect. Deep inside the topological phase, however, the localization is degraded by the vortices, as the decay constant $\xi^{-1}$ is considerably smaller than in the case without orbital magnetic field. This suggests that hybridization between Majorana and vortex states delocalizes the Majorana states. 
This is further corroborated by a comparison with the results in Fig.~\ref{fig:top}(b) and \ref{fig:coherence}, which were calculated using a stronger attractive interaction but otherwise unchanged parameters. There is almost no deviation between the induced coherence length at $y_c=0$ in the presence of vortices in Fig.~\ref{fig:coherence} (green curve) and the induced coherence length for a vortex-free superconductor in Fig.~\ref{fig:top}(b) (blue curve). The reason for this is that the coherence length in the superconducting substrate is short and the vortices have very little overlap with each other or with the wire. When the interaction strength is reduced, as in Fig.~\ref{fig:divided}, the vortex size grows and the Majorana decay length is enhanced by hybridization with vortex states even at $y_c=0$.

The increase of the induced coherence length due to vortices can be understood as a consequence of the hybridization between Caroli-de Gennes-Matricon states in different vortices, which leads to the formation of a band of subgap states in the superconducting substrate. The Majorana end states in the chain hybridize with the low-energy extended states in the vortex lattice which increases their localization length. Here, the extended nature of the vortex states is crucial, as a coupling to localized subgap states cannot lead to delocalization of the Majorana modes. In order to corroborate this interpretation, we have divided the superconductor into several isolated strips by eliminating hoppings $w^s$ along a series of cuts running perpendicular to the chain (see Fig.~\ref{fig:divided}). The coherence length shown in Fig.~\ref{fig:divided} monotonically decreases as the number of segments is increases while all other parameters, including the chain length, are kept the same. In the maximal case of $N_\text{segm}=20$ segments, when each segment has the length of a magnetic unit cell, the coherence length is comparable to the case without vortices over a large range of coupling strengths $t$. At the optimal value $t\simeq 0.6$, the segmentation of the superconductor leads to a three-fold reduction of the coherence length.

{\em Conclusions.}---
Sizable external magnetic fields are a requirement for some of the most promising topological superconductor realizations to generate one-dimensional helical liquids. While superconducting thin films are limited to field strengths of the order of $1-2\,$T, type-II superconductors can withstand much stronger fields exceeding $10\,$T \cite{Lee2017}. This increased durability comes at the expense of an inhomogeneous superconducting order parameter due to the presence of vortices. We have seen how phase gradients due to circulating supercurrents around the vortices as well as low-energy bound states in vortex cores can weaken localization of Majorana states and thus degrade their topological protection.
Our findings suggest that the consequences of the phase gradients can be diminished by locating the wire in the center between vortices. This could be reached by designing heterostructures with appropriate pinning or antipinning of vortices. To remedy the detrimental effect of vortex states, we suggest to divide the superconductor into multiple segments along the chain to avoid hybridization of Majorana modes with extended low-energy states. It would be interesting to extend this study to quasi one-dimensional chains, where orbital effects of the out-of-plane field also become important in the chain (cf.\ Ref.~\onlinecite{Nijholt2016}). Moreover, the effect of vortices is particularly relevant in the geometry of Ref.~\onlinecite{Lee2017}, where a thin superconducting strip is located between two coupled counterpropagating quantum Hall edge modes.

%

%

 \clearpage

\onecolumngrid
\begin{appendix}

\section{Self-consistency equation in the magnetic unit cell} \label{Appendix_A}

\begin{figure}[b] 
\begin{minipage}[h]{0.6\linewidth}
  \center{\includegraphics[width=1\linewidth]{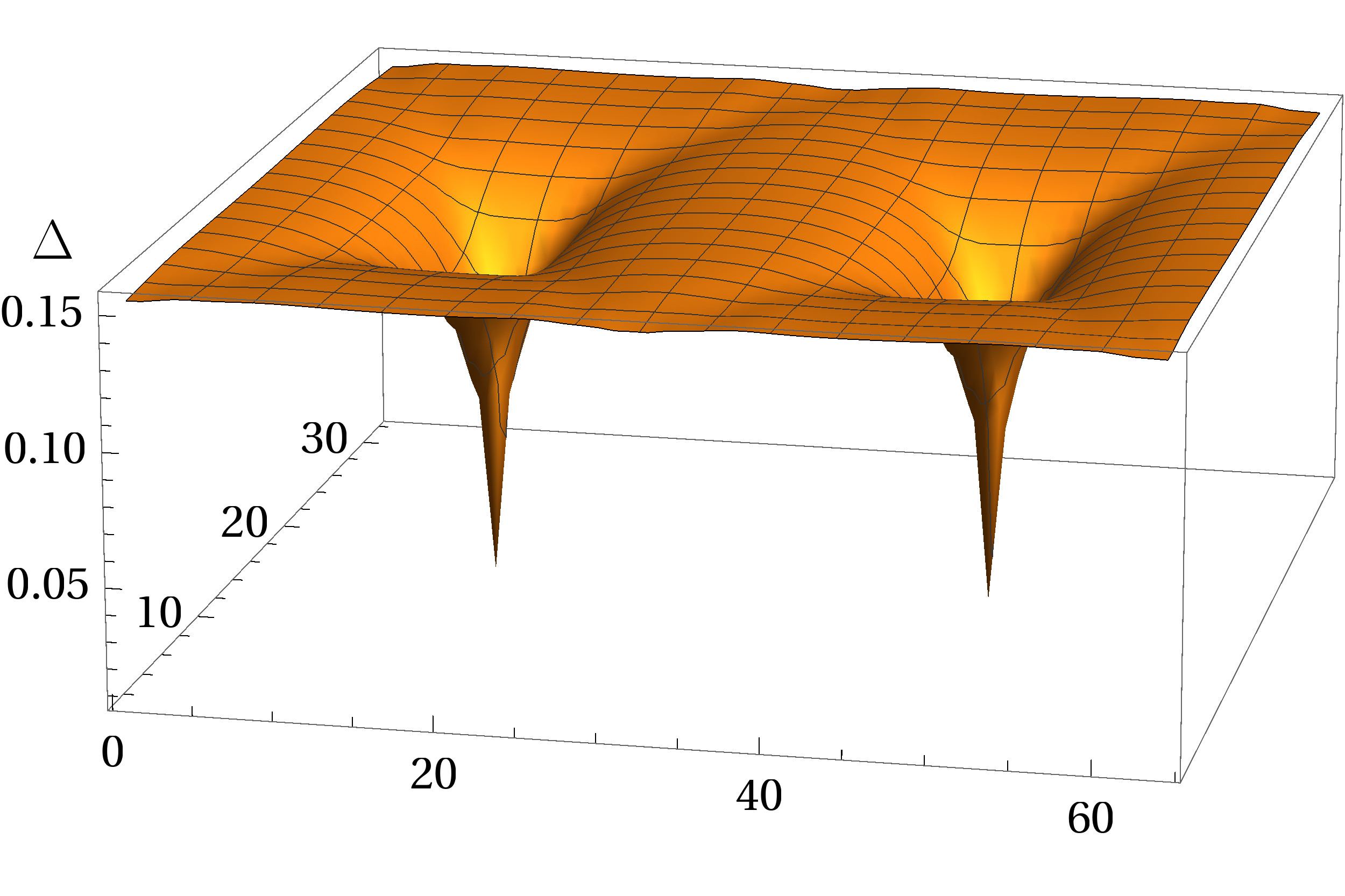}}
  
\end{minipage}  
  \caption{Spatial profile of $\Delta_i$ in a magnetic unit cell for $V=-1.6w^s$, $N_x\times N_y=64 \times 32$  }
  \label{fig:supp_delta}
\end{figure}

The mean-field superconductor Hamiltonian 
\begin{align}
\mathcal{H}_{mf}=\sum_i \Delta_i c^{\dag}_{i\uparrow} c^{\dag}_{i\downarrow}+\Delta^{*}_i c_{i\downarrow} c_{i\uparrow} ,\qquad 
\Delta_i=V\langle c_{i\downarrow}c_{i \uparrow}\rangle 
\end{align}
can be written in the Nambu basis $(c_\uparrow, c_\downarrow, c^{\dag}_\downarrow,-c^{\dag}_\uparrow)^T$ as
\begin{align}
\mathcal{H}_s =\begin{pmatrix} H_{ij} & 0 & \Delta & 0 \\ 0 & H_{ij} & 0 & \Delta \\ \Delta^* & 0 & -H^*_{ij} & 0\\ 0 & \Delta^* & 0 &-H^*_{ij}  \end{pmatrix} 
\end{align}
Due to the spin symmetry of the Hamiltonian, eigenstates at energy $E$ take the form
\begin{align}
\left(u(E), v^*(-E), v(E),-u^*(-E)\right)^T 
\end{align}
and the self-consistency relation reads
\begin{align}
\Delta_i=V \sum_E  v^*_i(E) u_i(E) f(E).
\end{align}
A typical spatial profile of the pairing strength $\Delta_i$ in a magnetic field is shown in Fig.~\ref{fig:supp_delta}.
The current that flows from site $i$ to $j$ can be derived from the continuity equation as
\begin{align}
 j_{i \rightarrow j}= \frac{i e t}{\hbar}\sum_n \left[(e^{-i \chi_{i j}}u_{n,i}u^{*}_{n,j}- e^{i \chi_{i j}}u^{*}_{n,i}u_{n,j})f(E_n)
+ (e^{-i \chi_{i j}}v^{*}_{n,i}v_{n,j}- e^{i \chi_{i j}}v_{n,i}v^{*}_{n,j})f(-E_n)\right]
 \end{align}
 where $\chi_{i j}=2\pi/\Phi_0 \int_i^j A(\vec{r})d\vec{r}$.\

The magnetic translational operators, which commute with Hamiltonian and themselves in the gauge $\vec{A}=-B y \vec{e}_x $ look as follows:
\begin{align}
\begin{split}
&(T_x^M)^{N_x}=\sum_{m,n}c^{\dag}_{m+N_x,n}c_{m,n}\\
&(T_y^M)^{N_y}=\sum_{m,n}c^{\dag}_{m,n+N_y}c_{m,n}e^{-\frac{2\pi m}{N_x}}.
\end{split}
\end{align}
The magnetic unit cell has the size $N_x \times N_y$, where $N_x N_y a^2 B = \Phi_0$. Given the translational operators, we can
formulate the magnetic Bloch theorem:
\begin{gather}
\begin{pmatrix}
u_k(\vec{r}+\vec{R}) \\
v_k(\vec{r}+\vec{R})
\end{pmatrix}
=e^{i \vec{k} \vec{R}}
\begin{pmatrix}
e^{i\chi(\vec{r},\vec{R})/2}u_k(\vec{r})\\
e^{-i\chi(\vec{r},\vec{R})/2}v_k(\vec{r})
\end{pmatrix},
\label{bloch}
\end{gather}
where $\chi(\vec{r},\vec{R})=4\pi n \, x/N_x$, $\vec{R}=m N_x \vec{e}_x+n N_y \vec{e}_y$, and
\begin{align}
\vec{k}=\frac{2\pi l_x}{N_x M_x}\vec{e}_x + \frac{2\pi l_y}{N_y M_y}\vec{e}_y, \quad l_i=0,1,..M_i-1, \quad i=\{x,y\}
\end{align}
The parameters $M_x$ and $M_y$ denote the number of unit cells in $x$ and $y$ directions. Now we can partially diagonalize the original Hamiltonian,
and perform the calculation for $M_x \times M_y$ matrices of $2 N_x N_y$ size, instead of one big matrix of $2 N_x M_x N_y M_y$ size.
The self-consistency equation changes in the following way:
\begin{align}
\Delta(\vec{r}+\vec{R})=&V\sum_{n}u_n(\vec{r}+\vec{R})v_n^*(\vec{r}+\vec{R})f(E_n)=\frac{V}{M_x M_y} e^{i\chi(\vec{r},\vec{R})}\sum_{n,k}u_{n,k}(\vec{r})v_{n,k}^*(\vec{r})f(E_{n,k})
\label{delta_1}\\
\Delta(\vec{r})=&\frac{V}{M_x M_y}\sum_{n,k}u_{n,k}(\vec{r})v_{n,k}^*(\vec{r})f(E_{n,k}).
\end{align}
With the help of the Bloch basis 
\begin{align}
\ket{\vec{r}_k,\mu}=\frac{1}{(M_x M_y)^{1/2}}\sum_{\vec{R}} e^{i \vec{k} \cdot \vec{R} +i\chi_{\mu}(\vec{r},\vec{R})/2} \ket{\vec{r}+\vec{R},\mu}
\label{bloch_basis}
 \end{align}
we can obtain real space Green function in the magnetic unit cells
 \begin{equation}
G(\vec{r}+\vec{R},\pvec{r}'+\pvec{R}',\mu,\nu)=\frac{1}{M_x M_y}\sum_{\vec{k}} e^{i \vec{k}\cdot (\vec{R}-\pvec{R}')} e^{i(\chi_{\mu}(\vec{r},\vec{R})-\chi_{\nu}(\pvec{r}',\pvec{R}'))/2}G(\vec{r},\pvec{r}',\mu,\nu,\vec{k})
\label{transformation}
\end{equation}

\section{Topological quantum numbers}\label{App_B}
In the presence of time-reversal symmetry, the topological $\mathbb{Z}$ number can be computed by off-diagonalizing the Hamiltonian in the particle-hole space using a unitary transformation
$$H=
\begin{pmatrix}
0& A(k) \\
 A^*(k) & 0 \\
\end{pmatrix}.
$$
The $\mathbb{Z}$ number is simply $W=\frac{1}{2\pi i}\int_0^{2\pi}\nabla_k\log{\det{A(k)}}dk$.
In magnetic field, time-reversal symmetry is broken by supercurrents. However, the particle-hole operator  $P=\Lambda K = i \sigma_y \tau_y K$ remains a symmetry. Therefore, it is possible to define a $\mathbb{Z}_2$ quantum number as
$$\mathbb{Z}_2=\text{sign}\{\text{Pf}(\Lambda H(E=0,k=0))\} \text{sign}\{\text{Pf}(\Lambda H(E=0,k=\pi))\}$$

\begin{figure}[tb] 
\begin{minipage}[h]{0.45\linewidth}
  \center{\includegraphics[width=1\linewidth]{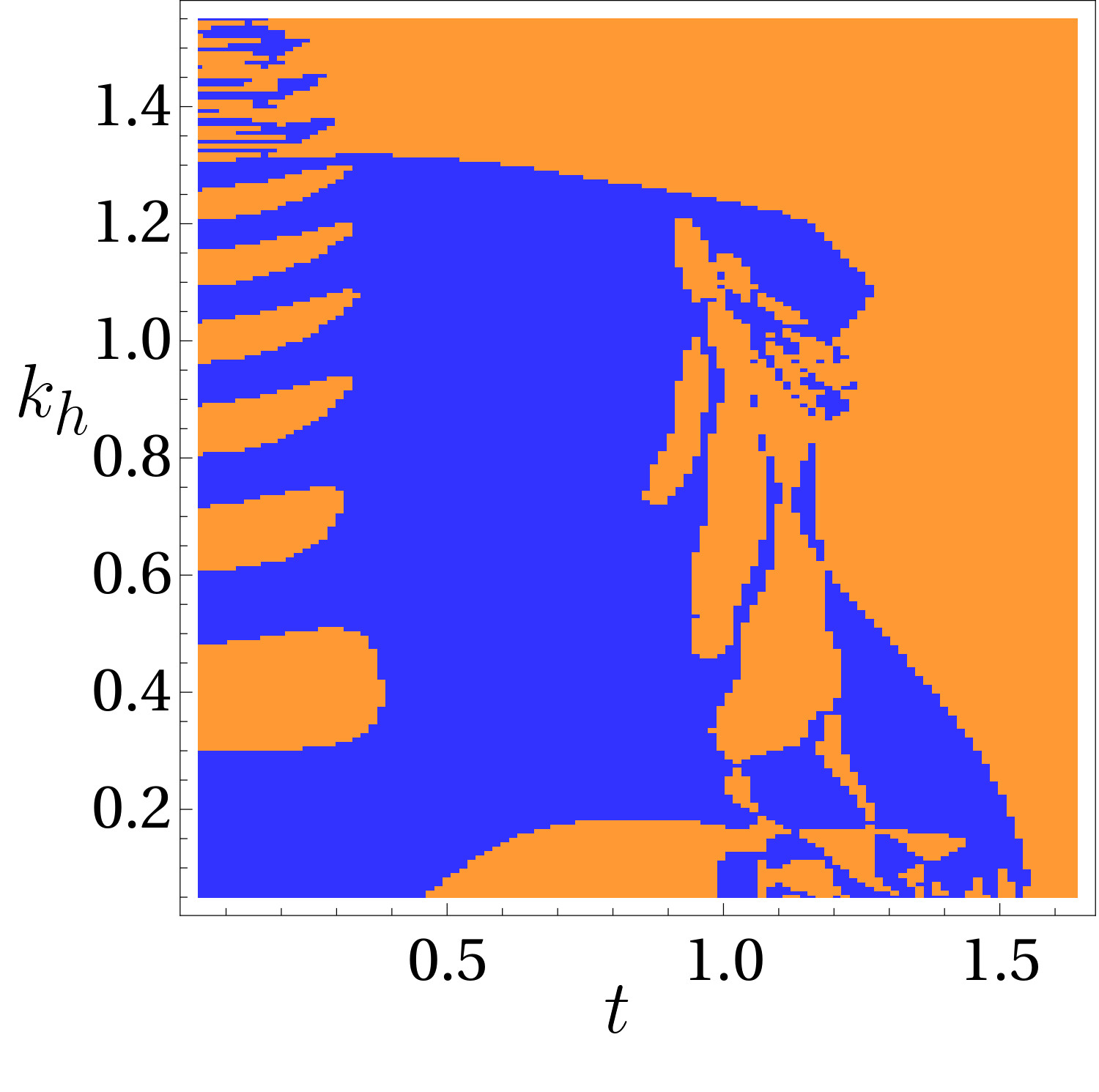}}
  \\Phase diagram for $y_c=2$
\end{minipage}  
  \hfill
\begin{minipage}[h]{0.45\linewidth}
  \center{\includegraphics[width=1\linewidth]{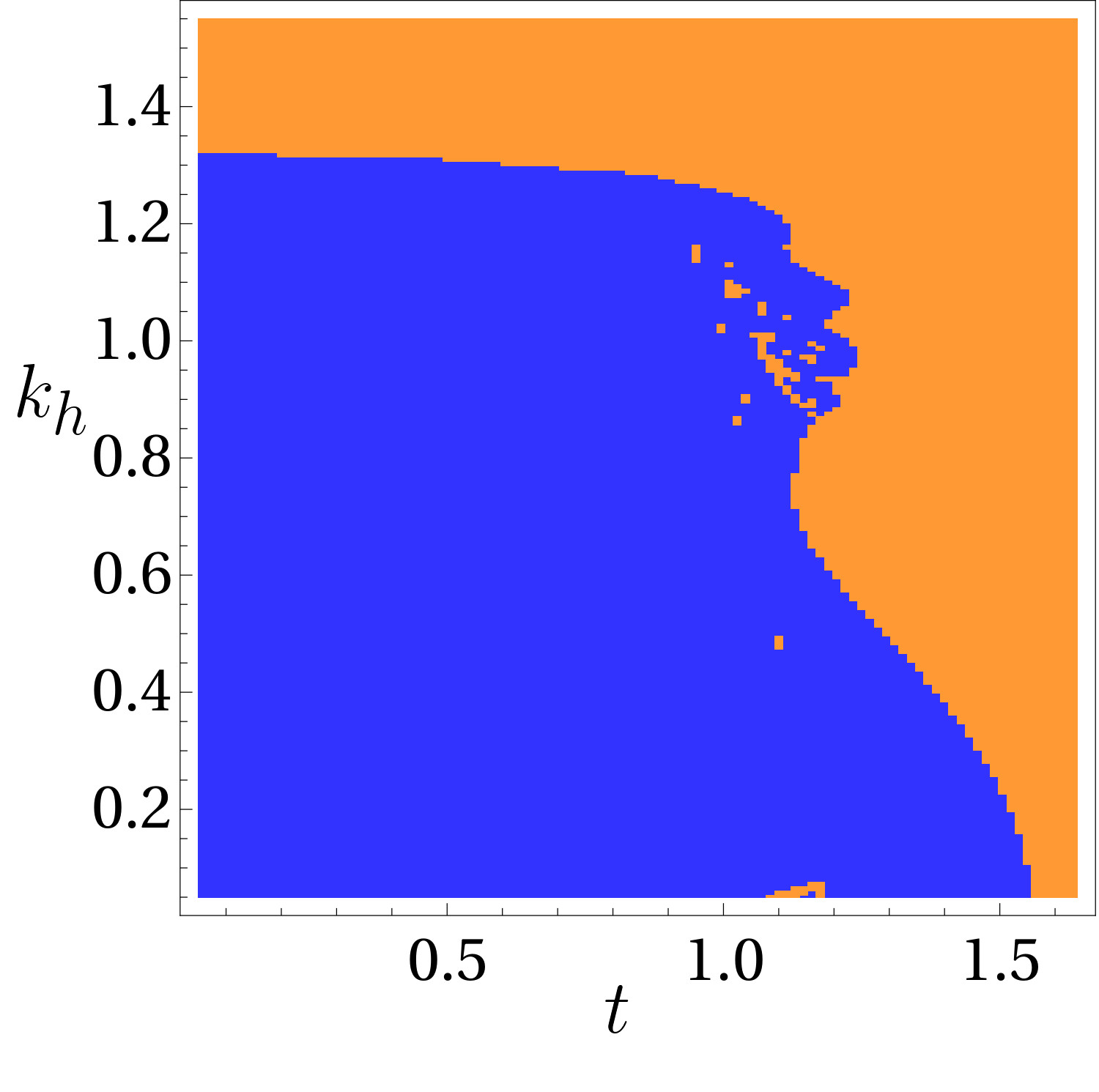}}
  \\ Phase diagram for $y_c=0$
\end{minipage}  
  \caption{The phase diagram for $y=0$ almost coincides with phase diagram for homegeneous superconductor, while case with $y=2$ differs significantly. Parameters are as in Fig.~\ref{fig:coherence} of the main text.}
  \label{fig:supp}
\end{figure}

In Fig.~\ref{fig:supp}, the $\mathbb{Z}_2$ phase diagrams for a wire placed at $y_c=0$ and $y_c=2$ are shown.

\end{appendix}

\end{document}